\documentclass[preprint,showpacs,showkeys,preprintnumbers,amsmath,amssymb]{revtex4}
 \usepackage{dcolumn}
 \usepackage{bm}

\begin{document}

 \title{ Hidden Conformal Symmetry of Self-Dual Warped AdS$_3$ Black Holes
 in Topological Massive Gravity }

 \author{ Ran Li }

 \thanks{Electronic mail: liran05@lzu.cn}

 \affiliation{Institute of Modern Physics,
 Chinese Academy of Sciences, Lanzhou, 730000, China}

 \author{ Ming-Fan Li }

 \thanks{Electronic mail: 11006139@zju.edu.cn}

 \affiliation{Zhejiang Institute of Modern Physics,
 Department of Physics,
 Zhejiang University, Hangzhou 310027, China}

 \author{ Ji-Rong Ren }

 \thanks{Electronic mail: renjr@lzu.edu.cn}

 \affiliation{Institute of Theoretical Physics,
 Lanzhou University, Lanzhou, 730000, China}

 \begin{abstract}

 We extend the recently proposal of
 hidden conformal symmetry to
 the self-dual warped AdS$_3$ black holes
 in topological massive gravity.
 It is shown that
 the wave equation of massive scalar field
 with sufficient small angular momentum
 can be reproduced by the
 SL(2, R) Casimir quadratic operator.
 Due to the periodic identification in the $\phi$ direction,
 it is found that only the left section of hidden conformal symmetry
 is broken to U(1), while the right section
 is unbroken, which only gives the left temperature
 of dual CFT.
 As a check of the dual CFT conjecture of self-warped AdS$_3$ black hole,
 we further compute
 the Bekenstein-Hawking entropy and
 absorption cross section and quasinormal modes of scalar
 field perturbation and show these are
 just of the forms predicted by the dual CFT.

 \end{abstract}



 \maketitle

 \section{Introduction}

 Topological massive gravity (TMG) is described by the theory
 of three dimensional Einstein gravity
 with a gravitational Chern-Simons
 correction and the cosmological constant \cite{TMG1,TMG2}.
 The well-known spacelike warped AdS$_3$ black hole
 \cite{warpedads} (previously obtained in \cite{clement}),
 which is a vacuum solution of topological massive gravity,
 is conjectured to be dual to a two dimensional
 conformal field theory (CFT) with non-zero left
 and right central charges \cite{TMGcentralcharge}.
 The spacelike warped AdS$_3$ black hole
 is a quotient of warped AdS$_3$ spacetime, just
 as BTZ black hole is a quotient of AdS$_3$
 spacetime. This leads to the breaking
 of the SL(2, R)$\times$SL(2, R) isometry
 of AdS$_3$ to the SL(2, R)$\times$U(1) isometry
 of warped AdS$_3$ black hole.
 It is shown in \cite{hiddenwarped} that,
 for a certain low energy limit,
 the wave equation of the massive scalar field in
 the background of spacelike warped AdS$_3$ black
 hole can be written as the Casimir operator of
 SL(2, R)$_L\times$SL(2, R)$_R$ Lie algebra,
 which uncovers the hidden SL(2, R)$\times$SL(2, R)
 symmetry of the wave equation of scalar field.

 Recently, a new class of solutions in
 three dimensional topological massive gravity
 named as self-dual warped AdS$_3$ black hole
 is proposed by Chen et al in \cite{chenselfdual}.
 It is conjectured that
 the self-dual warped AdS$_3$ black hole
 is dual to a chiral CFT with only
 nonvanishing left central charge,
 which is very different from the spacelike
 warped AdS$_3$ black hole.
 The self-dual warped AdS$_3$ black hole is
 locally equivalent to spacelike warped AdS$_3$
 spacetime via a coordinates transformation.
 The isometry group is just U(1)$_L\times$SL(2,R)$_R$,
 similar to the warped AdS$_3$ black hole.
 Under the consistent boundary condition,
 the U(1)$_L$ isometry is enhanced to a Virasoro
 algebra with nonvanishing left central charge,
 while the SL(2,R)$_R$ isometry becomes trivial
 with the vanishing right central charge, which
 is similar to the case of extremal Kerr/CFT
 correspondence \cite{GTSS,HMNS}. This suggests a novel example
 of warped AdS/CFT dual.

 In this paper, motivated by the recently proposed
 hidden conformal symmetry of the wave equation
 of scalar field propagating in the background of
 the general rotating black hole \cite{Castro},
 we consider the case
 of self-dual warped AdS$_3$ black holes in TMG.
 It is shown that
 the wave equation of massive scalar field
 propagating in the background of self-dual
 warped AdS$_3$ black hole
 can be rewritten in the form of
 Casimir operator of the
 SL(2, R)$_L\times$SL(2, R)$_R$ Lie algebra.
 Unlike the higher dimensional
 black holes where the near-horizon limit should be taken into
 account to match the wave equation with the Casimir operator,
 in the present case, only the condition of the small
 angular momentum of scalar field is imposed,
 which suggests that
 the hidden conformal symmetry
 is valid for the scalar field with arbitrary energy.
 So we have uncovered the hidden SL(2, R)$_L\times$SL(2, R)$_R$
 symmetry of the wave equation of massive scalar field
 in self-dual warped AdS$_3$ black hole.
 Then, we show that, due to the periodic identification
 in the $\phi$ direction,
 only one copy of hidden SL(2, R) symmetry
 is broken to U(1), while the another copy
 is unbroken. This only gives the left temperature $T_L$
 of dual CFT, while the right temperature $T_R$
 can not read from this approach. This point is also different
 from the higher dimensional black holes
 \cite{Castro,Krishnan,Chensun,Wang,
 chenlong,ranli,chendeyou,becker,
 chenlong1,chendeyou1,chenhuang,chenhuang1,addref}.
 Despite the right temperature
 can not be directly read from the
 periodic identification
 in the $\phi$ direction,
 one can still conjuncture that
 self-dual warped AdS$_3$ black hole
 is holographically dual to
 a two dimensional CFT
 with the left temperature $T_L=\frac{\alpha}{2\pi}$
 and the right temperature $T_R=\frac{x_+-x_-}{4\pi}$,
 which is exactly matches with
 the warped AdS/CFT correspondence suggested in \cite{chenselfdual}.

 As a check of this conjecture, we also
 show the entropy of the dual conformal field theory given by the
 Cardy formula matches exactly with the Bekenstein-Hawking entropy
 of self-dual warped AdS$_3$ black hole.
 Furthermore, the absorption cross section of scalar field
 perturbation calculated from the gravity side
 is in perfect match with that for a finite
 temperature 2D CFT. At last, we present an algebraic
 calculation of quasinormal modes for scalar field perturbation
 firstly proposed by Sachs et al in \cite{BTZTMG}.
 It is shown that the quasinormal modes
 coincide with the
 poles in the retarded Green's function
 obtained in \cite{chenselfdual},
 which is a prediction of AdS/CFT dual.

 This paper is organized as follows.
 In section II, we give a brief review of self-dual
 warped AdS$_3$ black hole in topological massive gravity.
 In section III,
 we study the hidden conformal
 symmetry of this black hole
 by analysing the wave equation of
 massive scalar field.
 In section IV,
 we give some interpretations of the dual conformal
 field description of self-warped AdS$_3$ black hole
 by computing the entropy, absorption cross section and
 quasinormal modes of scalar field and comparing
 the results from both gravity and CFT sides.
 The last section is devoted to discussion and conclusion.

 \section{Self-dual warped AdS$_3$ black hole}

 In this section, we will give a brief review of
 self-dual warped AdS$_3$ black hole in topological
 massive gravity.
 The action of topological massive gravity
 with a negative cosmological constant
 is given by
 \begin{eqnarray}
 I_{TMG}&=&\frac{1}{16\pi G}\int_\mathcal{M}
 d^3 x\sqrt{-g}\left(R+\frac{2}{l^2}\right)
 \nonumber\\&&
 +\frac{l}{96\pi G\nu}\int_\mathcal{M} d^3 x\sqrt{-g}\epsilon^{\lambda\mu\nu}
 \Gamma^{\alpha}_{\lambda\sigma}\left(\partial_{\mu}\Gamma^{\sigma}_{\alpha\nu}
 +\frac{2}{3}\Gamma^{\sigma}_{\mu\tau}\Gamma^{\tau}_{\nu\alpha}\right)\;.
 \end{eqnarray}
 Varying the above action with respect to
 the metric yields the equation of motion,
 which is given by
 \begin{eqnarray}
 G_{\mu\nu}-\frac{1}{l^2}g_{\mu\nu}+\frac{l}{3\nu}C_{\mu\nu}=0\;,
 \end{eqnarray}
 where $G_{\mu\nu}=R_{\mu\nu}-\frac{1}{2}Rg_{\mu\nu}$ is the
 Einstein tensor and $C_{\mu\nu}$ is the Cotton tensor
 \begin{eqnarray}
 C_{\mu\nu}=\epsilon_{\mu}^{\;\;\alpha\beta}\nabla_{\alpha}
 \left(R_{\beta\nu}-\frac{1}{4}R g_{\beta\nu}\right)\;.
 \end{eqnarray}

 Recently, a new class of solutions of
 topological massive gravity named
 as the self-dual warped AdS$_3$ black hole
 is investigated by Chen et al in \cite{chenselfdual}.
 The metric is given by
 \begin{eqnarray}
 ds^2&=&\frac{1}{\nu^2+3}\left(-\left(x-x_+\right)\left(x-x_-\right)d\tau^2
 +\frac{1}{\left(x-x_+\right)\left(x-x_-\right)}dx^2\right.\nonumber\\
 &&+\left.\frac{4\nu^2}{\nu^2+3}\left(\alpha d\phi+
 \frac{1}{2}\left(2x-x_+-x_-\right)d\tau\right)^2
 \right)\;,
 \end{eqnarray}
 where $x_+$ and $x_-$ are  the location of the
 outer and inner horizons respectively, and
 we have set $l=1$ for simplicity.
 The mass $M$ and angular momentum $J$ of this black hole
 are given by
 \begin{eqnarray}
 M=0\;,\;\;\;J=\frac{(\alpha^2-1)\nu}{6G(\nu^2+3)}\;.
 \end{eqnarray}
 The Hawking temperature $T_H$, angular velocity of
 the event horizon $\Omega_H$ and the
 Bekenstein-Hawking entropy $S_{BH}$ of
 this solution are respectively given by
 \begin{eqnarray}
 T_H&=&\frac{x_+-x_-}{4\pi}\;,\nonumber\\
 \Omega_H&=&-\frac{x_+-x_-}{2\alpha}\;,\nonumber\\
 S_{BH}&=&\frac{2\pi\alpha\nu}{3G(\nu^2+3)}\;.
 \end{eqnarray}

 This solution is asymptotic to the spacelike
 warped AdS$_3$ spacetime. It is shown in \cite{chenselfdual} that
 the self-dual warped AdS$_3$ black hole
 is locally equivalent to spacelike
 warped AdS$_3$ spacetime.
 Under coordinate transformation
 \begin{eqnarray}\label{eq7}
 v&=&\tan^{-1}\left[\frac{2\sqrt{(x-x_+)(x-x_-)}}{2x-x_+-x_-}
 \sinh\left(\frac{x_+-x_-}{2}\tau\right)\right]\;,\nonumber\\
 \sigma&=&\sinh^{-1}\left[\frac{2\sqrt{(x-x_+)(x-x_-)}}{2x-x_+-x_-}
 \cosh\left(\frac{x_+-x_-}{2}\tau\right)\right]\;,\nonumber\\
 u&=&\alpha\phi+\tan^{-1}\left[\frac{2x-x_+-x_-}{x_+-x_-}
 \coth\left(\frac{x_+-x_-}{2}\tau\right)\right]\;,
 \end{eqnarray}
 the metric of self-dual warped AdS$_3$ black hole solution
 can be transformed to the metric of spacelike warped
 AdS$_3$ spacetime
 \begin{eqnarray}
 ds^2=\frac{1}{\nu^2+3}\left(-\cosh^2\sigma dv^2+d\sigma^2
 +\frac{4\nu^2}{\nu^2+3}(du+\sinh\sigma dv)^2\right)\;.
 \end{eqnarray}
 It will be shown that this coordinates transformation
 is just the appropriate coordinates transformation
 to uncover the hidden conformal symmetry.
 The isometry group of this solution
 is U(1)$_L\times$SL(2,R)$_R$, which is generated by the killing
 vectors
 \begin{eqnarray}
 J_2=2\partial_u\;,
 \end{eqnarray}
 and
 \begin{eqnarray}
 \tilde{J}_1&=&2\sin v\tanh\sigma\partial_v
 -2\cos v\partial_\sigma+\frac{2\sin
 v}{\cosh\sigma}\partial_u\;,\nonumber\\
 \tilde{J}_2&=&-2\cos v\tanh\sigma\partial_v-2\sin v\partial_\sigma
 -\frac{2\cos v}{\cosh\sigma}\partial_u\;,\nonumber\\
 \tilde{J}_0&=&2\partial_v\;.
 \end{eqnarray}

 It is also shown in \cite{chenselfdual}
 that, under the consistent boundary condition,
 the U(1)$_L$ isometry is enhanced to a Virasoro
 algebra with the central charge
 \begin{eqnarray}\label{eq11}
 c_L=\frac{4\nu}{\nu^2+3}\;,
 \end{eqnarray}
 while the SL(2, R)$_R$ isometry becomes trivial
 with the vanishing central charge $c_R=0$,
 which is similar to the case of extremal Kerr/CFT
 dual \cite{GTSS,HMNS}. The entropy of self-dual warped AdS$_3$
 black hole can be reproduced by the
 Cardy formula. So it is conjectured that
 the self-dual warped AdS$_3$ black hole
 is holographically dual to a two dimensional
 chiral conformal field theory
 with nonvanishing left central charge.

 \section{Hidden conformal symmetry}

 In this section, we study the hidden
 conformal symmetry by analyzing the massive
 scalar field propagating in the background of
 self-dual warped AdS$_3$ black hole.
 Firstly, it is found that the scalar field
 equation can be exactly solved by the
 hypergeometric function. Then, by introducing the
 SL(2, R)$_L\times$SL(2, R)$_R$ generators and using
 the coordinates transformation (\ref{eq7}),
 we show the wave equation can be reproduced by
 the SL(2, R) Casimir operator.

 \subsection{Scalar field perturbation}

 Let us consider
 the scalar field $\Phi$ with mass $m$ in the background
 of self-dual warped AdS$_3$ black hole,
 where the wave equation is given by the
 Klein-Gordon equation
 \begin{eqnarray}
 \left(\frac{1}{\sqrt{-g}}\partial_\mu\left
 (\sqrt{-g}g^{\mu\nu}\partial_\nu\right)-m^2\right)\Phi=0\;.
 \end{eqnarray}
 The scalar field wave function $\Phi(\tau, x, \phi)$ can be expanded in
 eigenmodes as
 \begin{eqnarray}\label{eq13}
 \Phi=e^{-i\omega\tau+ik\phi}R(x)\;,
 \end{eqnarray}
 where $\omega$ and $k$ are the quantum numbers.
 Then the radial wave equation can be written as
 \begin{eqnarray}\label{eq14}
 \left[\partial_x\left((x-x_+)(x-x_-)\partial_x\right)
 +\frac{\left(\omega+\frac{x_+-x_-}{2\alpha}k\right)^2}{(x-x_+)(x_+-x_-)}
 -\frac{\left(\omega-\frac{x_+-x_-}{2\alpha}k\right)^2}{(x-x_-)(x_+-x_-)}\right]R(x)&&
 \nonumber\\
 =\left(-\frac{3(\nu^2-1)}{4\nu^2}\frac{k^2}{\alpha^2}+\frac{1}{\nu^2+3}m^2\right)
 R(x)&&.
 \end{eqnarray}

 This radial wave equation can be exactly solved by the
 hypergeometric function.
 In order to solve the radial equation,
 it is convenient to introduce the variable
 \begin{eqnarray}
 z=\frac{x-x_+}{x-x_-}\;.
 \end{eqnarray}
 Then, the radial equation can be rewritten in the form
 of hypergeometric equation
 \begin{eqnarray}
 z(1-z)\frac{d^2 R(z)}{dz^2}+
 (1-z)\frac{dR(z)}{dz}+\left(
 \frac{A}{z}+B+\frac{C}{1-z}\right)R(z)=0\;,
 \end{eqnarray}
 with the parameters
 \begin{eqnarray}
 A&=&\left(\frac{k}{2\alpha}+\frac{\omega}{x_+-x_-}\right)^2
 \;,\nonumber\\
 B&=&-\left(\frac{k}{2\alpha}-\frac{\omega}{x _+-x_-}\right)^2
 \;,\nonumber\\
 C&=&\frac{3(\nu^2-1)}{4\nu^2}\frac{k^2}{\alpha^2}-\frac{1}{\nu^2+3}m^2\;.
 \end{eqnarray}
 For later convenience, we consider the solution with the ingoing
 boundary condition at the horizon
 which is given by the hypergeometric
 function
 \begin{eqnarray}
 R(z)=z^{\alpha}(1-z)^{\beta}F(a,b,c,z)\;,
  \end{eqnarray}
 where
 \begin{eqnarray}
 \alpha=-i\sqrt{A}\;,\;\;\;
 \beta=\frac{1}{2}-\sqrt{\frac{1}{4}-C}\;,
 \end{eqnarray}
 and
 \begin{eqnarray}
 c&=&2\alpha+1\;,\nonumber\\
 a&=&\alpha+\beta+i\sqrt{-B}\;,\nonumber\\
 b&=&\alpha+\beta-i\sqrt{-B}\;.
 \end{eqnarray}

 It should be noted that, generally,
 the wave equation cannot be analytically solved and the
 solution must be obtained by matching solutions in an overlap region
 between the near-horizon and asymptotic regions.
 But, in the present case, we have
 shown that the radial equation
 can be exactly solved by hypergeometric functions.
 As hypergeometric functions
 transform in the representations of SL(2,R), this suggests the
 existence of a hidden conformal symmetry.
 In the next subsection, we will
 try to explore this hidden conformal symmetry.

 \subsection{SL(2, R)$_L\times$SL(2, R)$_R$}

 Now, we will uncover the hidden conformal symmetry
 by showing that the radial equation
 can also be obtained by using of the SL(2, R) Casimir operator.
 Let us define vector fields
 \begin{eqnarray}
 H_0&=&-\frac{i}{2}\tilde{J}_2\;,\nonumber\\
 H_1&=&\frac{i}{2}(\tilde{J}_0+\tilde{J}_1)\;,\nonumber\\
 H_{-1}&=&\frac{i}{2}(\tilde{J}_0-\tilde{J}_1)\;,
 \end{eqnarray}
 and
\begin{eqnarray}
 \tilde{H}_0&=&\frac{i}{2}J_2\;,\nonumber\\
 \tilde{H}_1&=&\frac{1}{2}(J_1+J_0)\;,\nonumber\\
 \tilde{H}_{-1}&=&\frac{1}{2}(J_1-J_0)\;,
 \end{eqnarray}
 with
 \begin{eqnarray}
 J_1&=&-\frac{2\sinh u}{\cosh\sigma}\partial_v
 -2\cosh u\partial_\sigma+2\tanh\sigma\sinh
 u\partial_u\;,\nonumber\\
 J_0&=&\frac{2\cosh u}{\cosh\sigma}\partial_v
 +2\sinh u\partial_\sigma-2\tanh\sigma\cosh u\partial_u\;.
 \end{eqnarray}
 Note that $(J_1, J_2, J_0)$ and
 $(\tilde{J}_1, \tilde{J}_2, \tilde{J}_0)$ are
 the SL(2, R)$_L\times$SL(2, R)$_R$ Killing vectors
 of AdS$_3$ spacetime.
 The vector fields $(H_1, H_0, H_{-1})$
 obey the SL(2, R) Lie algebra
 \begin{eqnarray}
 [H_0,H_{\pm 1}]=\pm iH_{\pm 1}\;,\;\;[H_{-1},H_1]=2iH_0\;,
 \end{eqnarray}
 and similarly for $(\tilde{H}_1, \tilde{H}_0, \tilde{H}_{-1})$.
 According to the coordinates transformation (\ref{eq7}),
 the SL(2, R) generators can be expressed
 in terms of the black hole coordinates $(\tau, x, \phi)$
 as
 \begin{eqnarray}\label{eq25}
 H_0&=&\frac{i}{2\pi T_R}\partial_\tau\;,\nonumber\\
 H_{-1}&=&ie^{-2\pi T_R\tau}\left[
 \sqrt{(x-x_+)(x-x_-)}\;\partial_x
 -\frac{1}{\sqrt{(x-x_+)(x-x_-)}}\cdot\frac{T_R}{T_L}\partial_\phi\right.
 \nonumber\\
 &&\;\;\;\;\;\;\;\;\;\;\;\;\;\;\;
 \left.+\frac{\left(x-\frac{x_++x_-}{2}\right)}{\sqrt{(x-x_+)(x-x_-)}}
 \cdot\frac{1}{2\pi T_R}\partial_\tau
 \right]\;,\nonumber\\
 H_{1}&=&ie^{2\pi T_R\tau}\left[
 -\sqrt{(x-x_+)(x-x_-)}\;\partial_x
 -\frac{1}{\sqrt{(x-x_+)(x-x_-)}}\cdot\frac{T_R}{T_L}\partial_\phi\right.
 \nonumber\\
 &&\;\;\;\;\;\;\;\;\;\;\;\;\;\;
 \left.+\frac{\left(x-\frac{x_++x_-}{2}\right)}{\sqrt{(x-x_+)(x-x_-)}}
 \cdot\frac{1}{2\pi T_R}\partial_\tau
 \right]\;,
 \end{eqnarray}
 and
 \begin{eqnarray}\label{eq26}
 \tilde{H}_0&=&\frac{i}{2\pi T_L}\partial_\phi\;,\nonumber\\
 \tilde{H}_{-1}&=&ie^{-2\pi T_L\phi}\left[
 \sqrt{(x-x_+)(x-x_-)}\;\partial_x
 +\frac{\left(x-\frac{x_++x_-}{2}\right)}{\sqrt{(x-x_+)(x-x_-)}}
 \cdot\frac{1}{2\pi T_L}\partial_\phi\right.\nonumber\\
 &&\;\;\;\;\;\;\;\;\;\;\;\;\;\;
 \left.-\frac{1}{\sqrt{(x-x_+)(x-x_-)}}\partial_\tau\right]\;,
 \nonumber\\
 \tilde{H}_{1}&=&ie^{2\pi T_L\phi}\left[
 -\sqrt{(x-x_+)(x-x_-)}\;\partial_x
 +\frac{\left(x-\frac{x_++x_-}{2}\right)}{\sqrt{(x-x_+)(x-x_-)}}
 \cdot\frac{1}{2\pi T_L}\partial_\phi\right.\nonumber\\
 &&\;\;\;\;\;\;\;\;\;\;\;\;\;\;
 \left.-\frac{1}{\sqrt{(x-x_+)(x-x_-)}}\partial_\tau\right]\;,
 \end{eqnarray}
 where $T_L$ and $T_R$ are defined by
 \begin{eqnarray}\label{eq27}
 T_L=\frac{\alpha}{2\pi}\;,\;\;\;
 T_R=\frac{x_+-x_-}{4\pi}\;.
 \end{eqnarray}

 The SL(2, R) quadratic Casimir operator is defined by
 \begin{eqnarray}
 \mathcal{H}^2=\mathcal{\tilde{H}}^2&=&-H_0^2+\frac{1}{2}(H_1H_{-1}+H_{-1}H_1)\;.
 \end{eqnarray}
 In terms of the $(\tau, x, \phi)$ coordinates,
 the SL(2, R) quadratic Casimir operator becomes
 \begin{eqnarray}
 \mathcal{H}^2&=&\partial_x\left((x-x_+)(x-x_-)\partial_x\right)
 -\frac{x_+-x_-}{x-x_+}\left[\frac{1}{4\pi T_R}\partial_\tau
 -\frac{1}{4\pi T_L}\partial_\phi\right]^2\nonumber\\
 &&+\frac{x_+-x_-}{x-x_-}\left[\frac{1}{4\pi T_R}\partial_\tau
 +\frac{1}{4\pi T_L}\partial_\phi\right]^2\;.
 \end{eqnarray}

 For the case of $\nu=1$, the first term of right hand side
 of Eq.(\ref{eq14}) is vanishing. It should be noted that,
 unlike the case of higher dimensional
 black holes \cite{Castro,Krishnan,Chensun,Wang,
 chenlong,ranli,chendeyou,becker,
 chenlong1,chendeyou1,chenhuang,chenhuang1,addref},
 where the near-region limit
 for the radial wave equation
 is taken into account,
 in the present case,
 no extra approximation is needed to
 match the wave equation of scalar field
 with the Casimir operator.
 So for the case of $\nu=1$,
 the self-dual warped black holes exhibit
 the local SL(2, R)$_L\times$SL(2, R)$_R$
 symmetry just like the BTZ black hole.

 We consider the nontrivial case of $\nu^2>1$,
 when these solutions are free of naked CTCs.
 Additional condition must be imposed to
 match the wave equation with the Casimir.
 In order to neglect the first term of right hand side
 of Eq.(\ref{eq14}), we impose the condition that
 the angular momentum $k$ of scalar field
 is sufficient small
 \begin{eqnarray}\label{eq29}
 \frac{3(\nu^2-1)}{4\nu^2}\frac{k^2}{\alpha^2}
 \ll 1\;.
 \end{eqnarray}
 Then, we find that the wave equation of massive scalar field
 with the sufficient small angular momentum $k$
 can be rewritten as the Casimir operator
 \begin{eqnarray}
 \mathcal{H}^2\Phi=\mathcal{\tilde{H}}^2\Phi
 =\frac{1}{\nu^2+3}m^2 \Phi\;.
 \end{eqnarray}
 and the conformal weights of dual
 operator of the massive scalar field $\Phi$
 is given by
 \begin{eqnarray}\label{eq32}
 (h_L,h_R)=\left(
 \frac{1}{2}+\sqrt{\frac{1}{4}+\frac{m^2}{\nu^2+3}},
 \frac{1}{2}+\sqrt{\frac{1}{4}+\frac{m^2}{\nu^2+3}}\right)\;.
 \end{eqnarray}

 So we have found that, similar to the case of higher dimensional
 black holes, the hidden SL(2, R)$_L\times$SL(2,R)$_R$
 symmetry of self-dual warped AdS$_3$ black hole
 is uncovered by investigating
 the wave equation of scalar field in its background.
 Note that the hidden conformal symmetry
 is not derived from
 the conformal symmetry of spacetime geometry itself.

 It is also interesting that the
 hidden conformal symmetry of self-dual warped AdS$_3$
 black hole is locally the isometry of AdS$_3$ spacetime,
 which means that scalar fields with sufficient small
 angular momentum $k$ satisfying the condition (\ref{eq29})
 do not feel the warped property of spacetime.
 While for the case of spacelike warped AdS$_3$ black hole
 investigated in \cite{hiddenwarped},
 this observation is valid for the
 scalar fields with sufficient low energy.

 \section{CFT interpretation}

 \subsection{Temperature and entropy}

 As pointed out in \cite{Castro},
 for the case of higher dimensional
 black hole, the vector fields
 of SL(2, R) generators are not globally defined.
 Because of the periodic identification
 in the $\phi$ direction, the hidden
 SL(2, R)$_L\times$SL(2,R)$_R$ symmetry
 is spontaneously broken to U(1)$_L\times$U(1)$_R$ subgroup,
 which gives rise to the left and right
 temperatures of dual CFT.

 The story is somewhat different
 in the present case.
 The generators of SL(2, R) presented in
 Eq.(\ref{eq26}) are affected by the periodic identification
 in the $\phi$ direction, while these in
 Eq.(\ref{eq25}) are not because they are just
 the Killing vectors associated to SL(2, R)
 isometry of this solution.
 This means that
 only one copy of hidden conformal symmetry
 is broken to U(1), while the another copy
 is unbroken, which only gives the left temperature $T_L$
 of dual CFT. The right temperature $T_R$
 can not read from this approach.
 The periodical identification
 makes no contribution to the right temperature.
 This point can also be reached from
 the coordinates transformation (\ref{eq7})
 which indicates that the self-dual
 warped AdS$_3$ black hole can not be obtained
 as a quotient of warped AdS$_3$ vaccum.

 As discussed in \cite{chenselfdual},
 the left and right temperatures
 of dual CFT can be defined
 with respect to the Frolov-Thorne vacuum \cite{frolov}.
 Consider the quantum field with
 eigenmodes of the asymptotic
 energy $\omega$ and angular momentum
 $k$. After tracing over the region inside
 the horizon, the vacuum is a diagonal
 density matrix in the energy-angular
 momentum eigenbasis with a Boltzmann weighting
 factor $e^{-\frac{\omega-k\Omega}{T_H}}$.
 The left and right charges $\omega_L$, $\omega_R$ associated
 to $\partial_\phi$ and $\partial_t$ are
 $k$ and $\omega$ respectively.
 In terms of these variables, the Boltzmann factor is
 \begin{eqnarray}
 e^{-\frac{\omega-k\Omega}{T_H}}=
 e^{-\frac{\omega_L}{T_L}-\frac{\omega_R}{T_R}}\;,
 \end{eqnarray}
 which gives the definition of left and right temperatures
 (\ref{eq27}).

 So one can conjecture that the
 self-dual warped AdS$_3$ black hole
 is holographically dual to
 a two dimensional CFT
 with the left temperature $T_L=\frac{\alpha}{2\pi}$
 and the right temperature $T_R=\frac{x_+-x_-}{4\pi}$.
 As a check of this conjecture,
 we now want to calculate the
 microscopic entropy of the dual CFT,
 and compare it with the Bekenstein-Hawking entropy
 of self-dual warped AdS$_3$ black hole.
 By imposing the consistent boundary condition,
 Chen et al \cite{chenselfdual}
 have calculated the central charge
 of the asymptotic symmetry group,
 where the conclusion is presented in Eq.(\ref{eq11}).
 So the microscopic entropy of the dual conformal field
 can be calculated by using the Cardy formula
 \begin{eqnarray}
 S_{CFT}=\frac{\pi^2}{3}(c_LT_L+c_RT_R)
 =\frac{2\pi\alpha\nu}{3G(\nu^2+3)}=S_{BH}\;,
 \end{eqnarray}
 which matches with the Bekenstein-Hawking entropy
 of self-dual warped AdS$_3$ black hole.

 \subsection{Absorption cross section}

 In this subsection, we will calculate the
 absorption probability for the scalar filed
 perturbation and compare it with result from the
 CFT side. For the spacelike warped AdS$_3$
 black hole, this aspect has been investigated in
 \cite{kim} and \cite{wwyu}.

 Under the condition (\ref{eq29}), the solution
 to the radial equation of scalar field perturbation with the ingoing
 boundary condition is explicitly given by
 \begin{eqnarray}
 R(x)&=&\left(\frac{x-x_+}{x-x_-}\right)
 ^{-i\left(\frac{k}{2\alpha}+\frac{\omega}{x_+-x_-}\right)}
 \left(\frac{x_+-x_-}{x-x_-}\right)^{\frac{1}{2}-\sqrt{\frac{1}{4}+\frac{m^2}{\nu^2+3}}}\nonumber\\
 &&\times F\left(\frac{1}{2}-\sqrt{\frac{1}{4}+\frac{m^2}{\nu^2+3}}
 -i\frac{2\omega}{x_+-x_-},\frac{1}{2}-\sqrt{\frac{1}{4}+\frac{m^2}{\nu^2+3}}
 -i\frac{k}{\alpha},\right.\nonumber\\
 &&\left.1-i\left(\frac{k}{\alpha}+\frac{2\omega}{x_+-x_-}\right),\;\;
 \frac{x-x_+}{x-x_-}\right)\;.
 \end{eqnarray}
 At asymptotic infinity, the solution
 behaves as
 \begin{eqnarray}
 R(x\rightarrow\infty)\sim
 Ax^{-\frac{1}{2}+\sqrt{\frac{1}{4}+\frac{m^2}{\nu^2+3}}}\;,
 \end{eqnarray}
 with
 \begin{eqnarray}
 A=\frac{\Gamma\left(1-i\left(\frac{k}{\alpha}+
 \frac{2\omega}{x_+-x_-}\right)\right)
 \Gamma\left(2\sqrt{\frac{1}{4}+\frac{m^2}{\nu^2+3}}\right)}
 {\Gamma\left(\frac{1}{2}+\sqrt{\frac{1}{4}
 +\frac{m^2}{\nu^2+3}}-i\frac{2\omega}{x_+-x_-}\right)
 \Gamma\left(\frac{1}{2}+\sqrt{\frac{1}{4}
 +\frac{m^2}{\nu^2+3}}-i\frac{k}{\alpha}\right)}\;.
 \end{eqnarray}
 The absorption cross section is then proportional to
 \begin{eqnarray}
 P_{\textrm{abs}}&\sim&\left|A\right|^{-2}\nonumber\\
 &\sim&\sinh\left(\frac{\pi k}{\alpha}+
 \frac{2\pi\omega}{x_+-x_-}\right)
 \left|\Gamma\left(\frac{1}{2}+\sqrt{\frac{1}{4}
 +\frac{m^2}{\nu^2+3}}-i\frac{2\omega}{x_+-x_-}\right)
 \right|^2\nonumber\\&&\times\left|
 \Gamma\left(\frac{1}{2}+\sqrt{\frac{1}{4}
 +\frac{m^2}{\nu^2+3}}-i\frac{k}{\alpha}\right)
 \right|^2\;.
 \end{eqnarray}

 To compare it with the result from the
 CFT side, we need to find out
 the related parameters.
 From the first law of black hole thermodynamics
 \begin{eqnarray}
 \delta S_{BH}=\frac{\delta M-\Omega_{H}\delta J}{T_{H}}\;,
 \end{eqnarray}
 one can calculate the conjugate charges
 \begin{eqnarray}
 \delta S_{BH}=\frac{\delta E_L}{T_L}
 +\frac{\delta E_R}{T_R}\;.
 \end{eqnarray}
 The solution is given by
 \begin{eqnarray}
 \delta E_L=\delta J\;,\;\;\delta E_R=\delta M\;.
 \end{eqnarray}
 Identifying $\delta M=\omega$ and $\delta J=k$,
 one can find the left and right conjugate charges
 as
 \begin{eqnarray}
 \omega_L\equiv\delta E_L=k\;,\;\;\omega_R\equiv\delta E_R=\omega\;,
 \end{eqnarray}
 which are coincide with the the left and right charges
 given in the last subsection.
 Finally, the absorption cross section can be expressed as
 \begin{eqnarray}
 P_{\textrm{abs}}\sim T_L^{2h_L-1}T_R^{2h_R-1}
 \sinh\left(\frac{\omega_L}{2T_L}+\frac{\omega_R}{2T_R}\right)
 \left|\Gamma\left(h_L+i\frac{\omega_L}{2\pi T_L}\right)\right|^2
 \left|\Gamma\left(h_R+i\frac{\omega_R}{2\pi T_R}\right)\right|^2\;,
 \end{eqnarray}
 which is precisely coincide with the
 the absorption cross section for a finite
 temperature 2D CFT.

 \subsection{Quasinormal modes}

 In this subsection, we will compute
 the quasinormal modes of scalar field perturbation
 by using the algebraic method firstly
 proposed by Sachs et al in \cite{BTZTMG}
 and compare the results with that presented in
 \cite{ranlimode}.
 This method strongly depends on the
 observation of hidden conformal symmetry
 in the last section.
 This method has also been employed to
 investigate the quasinormal modes of
 vector and tensor perturbation
 by chen et al in \cite{chenbinmode}.

 Here, we consider the general case without
 imposing the small angular momentum condition
 (\ref{eq29}). The radial equation of scalar field
 perturbation can be written using the SL(2, R) generators
 as
 \begin{eqnarray}\label{eq44}
 \left[\frac{1}{2}\left(H_1H_{-1}+H_{-1}H_1\right)-H_0^2
 +\frac{3(\nu^2-1)}{4\nu^2}\tilde{H}_0^2
 \right]\Phi=\frac{m^2}{\nu^2+3}\Phi\;.
 \end{eqnarray}
 Firstly, we consider the chiral highest weight modes satisfying the condition
 \begin{eqnarray}
 H_1\Phi=0\;.
 \end{eqnarray}
 Under the ansatz (\ref{eq13}) for scalar field,
 this condition implies the equation
 \begin{eqnarray}
 \frac{d}{dx}\ln R(x)=-\frac{i}{(x-x_+)(x-x_-)}
 \left(\frac{T_R}{T_L}k+\frac{[(x-x_+)+(x-x_-)]}{4\pi T_R}\omega
 \right)\;,
 \end{eqnarray}
 which gives the solution
 \begin{eqnarray}
 R(x)=(x-x_+)^{-i\left(\frac{\omega}{4\pi T_R}+\frac{k}{4\pi T_L}\right)}
 (x-x_-)^{-i\left(\frac{\omega}{4\pi T_R}-\frac{k}{4\pi
 T_L}\right)}\;.
 \end{eqnarray}
 Then the operator equation (\ref{eq44}) can be transformed into
 an algebra equation
 \begin{eqnarray}
 \left(\frac{\omega}{2\pi T_R}\right)^2
 +i\left(\frac{\omega}{2\pi T_R}\right)-C=0\;.
 \end{eqnarray}
 The solution of this equation gives
 the lowest quasinormal mode in the right-moving sector
 \begin{eqnarray}
 \frac{\omega}{2\pi T_R}=-i h'_R\;,\;\;
 h'_R=\frac{1}{2}+\sqrt{\frac{1}{4}-C}\;.
 \end{eqnarray}
 It should be noted that this conformal
 weight is slightly different from that
 given by Eq.(\ref{eq32}). If the small angular
 momentum limit is taken into account,
 it recovers the conformal weight for scalar field
 given by Eq.(\ref{eq32}).
 Then, the descendents of the chiral highest weight mode $\Phi$
 \begin{eqnarray}
 \Phi^{(n)}=(H_{-1}\tilde{H}_{-1})^{n}\Phi\;,
 \end{eqnarray}
 give the infinite tower of quasinormal modes
 \begin{eqnarray}
 \frac{\omega}{2\pi T_R}=-i(n+h'_R)\;.
 \end{eqnarray}
 This is just the result obtained in \cite{ranlimode,chenbinmode},
 and coincides with the
 poles in the retarded Green's function
 obtained in \cite{chenselfdual}.

 The left sector of quasinormal scalar modes
 can not be obtained analogously.
 The solution constructed
 from the chiral highest weight condition
 $H_1\Phi=0$ falls off in time as well as at infinity,
 while for another highest weight condition $\tilde{H}_1\Phi=0$
 one can not obtain the solution with the same property.

 \section{Conclusion}

 We have investigated the hidden conformal symmetry
 of self-dual warped AdS$_3$ black holes in topological
 massive gravity.
 The wave equation of massive scalar field
 propagating in this background
 with sufficient small angular momentum
 can be rewritten in the form of
 SL(2, R) Casimir operator.
 Interestingly, unlike the higher dimensional
 black holes where the near-horizon limit should be taken into
 account to match the wave equation with the Casimir operator,
 in the present case, only the condition of the small
 angular momentum of scalar field is imposed,
 which suggests that
 the hidden conformal symmetry
 is valid for the scalar field with arbitrary energy.

 Despite the right temperature
 can not be directly read from the
 periodic identification
 in the $\phi$ direction,
 one can still conjecture that the
 self-dual warped AdS$_3$ black hole
 is dual to
 a 2D CFT
 with nonzero left and right temperatures.
 As a check of this conjecture, we also
 show that the entropy of the dual conformal field given by the
 Cardy formula matches exactly with the Bekenstein-Hawking entropy
 of self-dual warped AdS$_3$ black hole.
 Furthermore, the absorption cross section of scalar field
 perturbation calculated from the gravity side
 is in perfect match with that for a finite
 temperature 2D CFT. At last, an algebraic
 calculation of quasinormal modes for scalar field perturbation
 is presented and the correspondence between quasinormal modes
 and the poles in the retarded Green's function
 is found.

 \section*{Acknowledgement}

 RL would like to thank Shi-Xiong Song and Lin-Yu Jia for
 helpful discussions.
 The work of JRR was supported by the Cuiying Programme of Lanzhou
 University (225000-582404) and the Fundamental Research Fund for
 Physics and Mathematic of Lanzhou University(LZULL200911).

 \end{document}